\documentclass[prd,preprint,superscriptaddress,amsmath,amssymb,nofootinbib]{revtex4}
\usepackage{graphicx}
\usepackage{dcolumn}
\usepackage{bm}
\usepackage{amssymb}
\usepackage{amsmath}
\usepackage{epsfig}    
\usepackage{color}
\usepackage{slashed}
\usepackage{hhline}
\usepackage{bbm}
\usepackage{tikz-feynman}
\tikzfeynmanset{compat=1.1.0}

\def\be{\begin{equation}}
\def\ee{\end{equation}}
\newcommand{\bea}{\begin{eqnarray}}
\newcommand{\eea}{\end{eqnarray}}
\newcommand{\nn}{\nonumber}

\begin{document}

\title{A theoretical account of tiny multi-Higgs vacuum expectation values from non-invertible symmetry}

\author{Takaaki Nomura}
\email{nomura@scu.edu.cn}
\affiliation{College of Physics, Sichuan University, Chengdu 610065, China}

\author{Hiroshi Okada}
\email{hiroshi3okada@htu.edu.cn}
\affiliation{Department of Physics, Henan Normal University, Xinxiang 453007, China}
 
\date{\today}

\begin{abstract}
We propose a novel mechanism to explain the naturally small vacuum expectation values (VEVs) of exotic multi-Higgs fields by employing non-invertible symmetries. Specifically, we introduce an $SU(2)_L$ quartet $H_4$ and a quintet $H_5$ within the framework of the minimal Fibonacci fusion rule (FFR). This non-invertible symmetry strictly forbids the generation of tree-level VEVs for these exotic fields. However, once the symmetry is broken, these VEVs are generated radiatively at the one-loop level.
This mechanism is highly minimal, as it requires no additional loop-inducing particles. 
We demonstrate that for various benchmark energy scales, the resulting VEVs are naturally suppressed to the order of $10^{-3}-10^{-2}$ GeV, 
satisfying experimental constraints from the $\rho$ parameter.
Finally, we illustrate the phenomenological viability of this setup by applying it to three representative neutrino mass models: the type-III seesaw, the Dirac neutrino seesaw, and the inverse seesaw mechanisms.
\end{abstract} 
\maketitle
\newpage

\section{Introduction}
The idea of introducing large Higgs multiplets beyond the Standard Model (SM) doublet has long been recognized as a fruitful direction in model building, particularly in neutrino mass models and related frameworks. The essential feature is that the VEVs of such exotic Higgs fields are desired to be naturally small, thereby allowing the couplings to remain within experimentally accessible ranges. In other words, we want these constructions to provide highly testable models.
A generic consequence of introducing such an exotic Higgs multiplet is a deviation in the so‑called rho parameter, defined as
\begin{align}
\rho=\frac{m^2_W}{m_Z^2 \cos^2\theta_W},
\end{align}
 where $m_W$ is the charged-gauge boson mass, $m_Z$ is the neutral gauge boson mass in the SM, and $\sin\theta_W$ is the Weinberg angle.
Experimentally, this quantity is nearly unity, with deviations below $2.5\times10^{-4}$~\cite{ParticleDataGroup:2024cfk}.
 To satisfy this stringent bound, the VEVs of additional large Higgs multiplets must typically be below the GeV scale—roughly two orders of magnitude smaller than the SM Higgs VEV,  $v_H=246$ GeV.
 Thus, the smallness of multi‑Higgs VEVs is not merely an assumption but is experimentally guaranteed through precision measurements of the rho parameter.\\
From a theoretical perspective, however, explaining the smallness of these VEVs remains a significant challenge.
 In particular, it is desirable to realize small VEVs while maintaining all the scalar masses at the electroweak to TeV scale from the viewpoint of testability.
Several representative ideas have been proposed to address this.
For instance, type‑II seesaw models can generate small VEVs radiatively at the one-loop level~\cite{Kanemura:2012rj}, and two‑Higgs‑doublet models (THDMs) with loop‑induced VEVs have also been explored~\cite{Kanemura:2011mw}. 
However, many of these constructions rely on introducing inert bosons or new group symmetries such as $U(1)_{B-L}$.
While such extensions are phenomenologically rich—leading to dark matter candidates or new gauge bosons—they are not minimal if the primary goal is to suppress VEVs.

In this work, we propose a novel mechanism to realize small VEVs for the exotic scalar quadruplet $H_4$ and quintuplet $H_5$ by employing the Fibonacci fusion rule (FFR), which is recognized as the minimal non‑invertible symmetry and has recently been explored in phenomenological contexts~\cite{Okada:2026pek, Xu:2026nwh, Okada:2026bpp, Nomura:2026hli, Qu:2026omn, Kobayashi:2025ldi, Kobayashi:2024cvp, Kobayashi:2025znw,
Nomura:2025yoa, Suzuki:2025oov, Kobayashi:2025cwx, Nomura:2025sod,  Okada:2025kfm, Jangid:2025krp, Jangid:2025thp, Nomura:2025tvz, Okada:2025adm, Okada:2026gxl, Chen:2025awz, Okada:2026iob, Liang:2025dkm,Kobayashi:2025thd, Kobayashi:2025rpx, Nakai:2025thw, Kobayashi:2025lar, Kobayashi:2025wty, Heckman:2024obe,Kaidi:2024wio,Funakoshi:2024uvy}.  
It consists of two commutable algebras $({\mathbbm I},\tau)$ with the following multiplication rules:  
\begin{align}
\tau\otimes\tau={\mathbbm I}\oplus \tau,\quad {\mathbbm I}\otimes \tau=\tau,\quad {\mathbbm I}\otimes {\mathbbm I}={\mathbbm I}.
\end{align}
Here, $H_4$ denotes an $SU(2)_L$ quadruplet with hypercharge $Y=1/2$, while $H_5$ represents an $SU(2)_L$ quintuplet with hypercharge $Y=1$. 
A key feature of the non‑invertible symmetries is that they forbid tree‑level VEVs, while still allowing them to be radiatively generated at the one-loop level once the symmetry is  dynamically broken. 
Crucially, this mechanism does not require the introduction of additional loop‑inducing particles, making it more minimal than conventional approaches. 
We then present three representative applications of these tiny VEVs: type-III seesaw, Dirac seesaw, and inverse seesaw. 
In each case, we consider the renormalization group evolution (RGE) of the SU(2)$_L$ gauge coupling $g_2$ to identify the valid energy scales for respective models.

This paper is organized as follows.
In Section II, we review the mechanism for generating tiny VEVs in large multiplet Higgs fields.
In section III, we present three illustrative applications, discussing the RGE for each model and determining the corresponding cut-off energy scales. 
Finally, in section IV, we  summarize our findings and provide concluding remarks.

\section{Framework}
\begin{figure}
\centering
\begin{tikzpicture}
  \node[ellipse, draw, dashed,
        minimum width=2.5cm, minimum height=2.5cm] (cluster) {};
  \node (lambda1) at (cluster.west) [xshift=15pt] {$\lambda_{H_2 H_4}^{(j)}$};
  \node (lambda2) at (cluster.east) [xshift=-10pt] {$\lambda_{0}^{(i)}$};  
  \node (phi1) at ($(lambda1)+(-1.8cm,1.2cm)$) {$H_2^\dag$};
  \node (phi2) at ($(lambda1)+(-1.8cm,-1.2cm)$) {$H_2$};
  \draw[dashed] ($(lambda1)-(15pt,-1pt)$) -- (phi1);
  \draw[dashed] ($(lambda1)-(15pt,-1pt)$) -- (phi2);
  \node (phi3) at (cluster.north) [yshift=8pt] {$H^\dag_4$};
  \node (phi4) at (cluster.south) [yshift=-8pt] {$H_4$};
  \draw[dashed] (phi3) -- (cluster.north);
  \draw[dashed] (phi4) -- (cluster.south);
  \node (phi5) at ($(lambda2)+(1.8cm,1.2cm)$) {$H_4^*$};
  \node (phi6) at ($(lambda2)+(1.8cm,-1.2cm)$) {$H_2^T$};
  \draw[dashed] ($(lambda2)+(9.5pt,0cm)$) -- (phi5);
  \draw[dashed] ($(lambda2)+(9.5pt,0cm)$) -- (phi6);
\end{tikzpicture}
\caption{One-loop diagram of  the quartic term $\delta\lambda_{ij}$.}
\label{fig:1lp_higgs}
\end{figure}
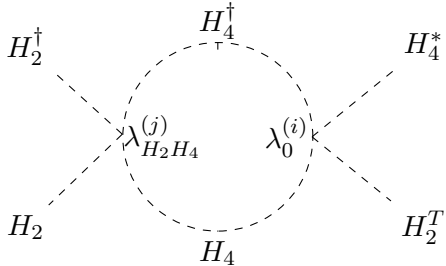

In this section, we provide a detailed review of our framework showing how the tiny VEVs are generated under the FFR.
By assigning the FFR algebra $\tau$ to the exotic Higgs fields $H_4$ and $H_5$, while keeping the SM Higgs $H$
 neutral, we construct the scalar potential $\mathcal{V}\equiv \mathcal{V}_2+\mathcal{V}_3+\mathcal{V}_4$ as follows:
 \begin{align}
\mathcal{V}_2 &= -\mu_{H_2}^2 H_2^\dagger H_2 + \mu_4^2 H_4^\dagger H_4 + \mu_5^2 H_5^\dagger H_5,\\
\mathcal{V}_3 &= -\mu_0 H_2H_5^\dagger H_4 +{\rm c.c.}, \\
 \mathcal{V}_4&=\sum_{i=1,2} \lambda_0^{(i)} [H_4^\dag H_2 H_4^\dag H_4]_i 
+\sum_{i=1,2} \lambda_{H_2H_4}^{(i)} [H_2^\dag H_2 H_4^\dag  H_4]_i +{\rm c.c.}
+V_4^{\rm trivial}
 \label{eq:pot4-1},
\end{align}
where we assume that $\mathcal{V}_2$ effectively corresponds to the mass  terms for each multiplets,
and $V_4^{\rm trivial}$ is the trivial quartic terms such as $\sum _{a\le b=2,4,5}\lambda_{H_a H_b} |H_a|^2|H_b|^2$.
$[\cdots]$ denotes all the combinations of trivial singlets under the $SU(2)_L$.
 The mass parameters $\mu_{0,4,5}$ are considered to be TeV scale from the viewpoint of testability in future experiments. Then, as we choose positive $\mu_4^2$ and $\mu_5^2$, the minimum of the scalar potential leads to $\langle H_4 \rangle = \langle H_5 \rangle =0$ at tree level.
The VEV of $H_4$ can be induced if we have the term $[H^\dag_2 H^*_4 H^T_2 H_2]$ that is forbidden at tree level due to the FFR~\cite{Xu:2026nwh}.
However, this term can be generated through $ \lambda_{H_2H_4}^{(i)} $ and $ \lambda_0^{(i)} $ interactions at one-loop level as shown in Fig.~\ref{fig:1lp_higgs}. 
Clearly, the FFR is dynamically violated at the one-loop level,  which is a characteristic feature of non-invertible symmetries. The quartic coupling of $[H^\dag_2 H^*_4 H^T_2 H_2]_{ij}$ is computed as 
\begin{align}
\delta\lambda_{ij} \approx 
- \frac{\lambda_0^{(i)} \lambda_{H_2 H_4}^{(j)}}{(4\pi)^2}
\left[
\frac{2-2r_4+(1+r_4)\ln(r_4)}{1-r_4}
\right],
\end{align}
where $r_4\equiv \mu_{4}^2/\Lambda^2$, with $\Lambda$ being the cut-off scale, which we will discuss later. 
%

Solving the stationary conditions, $\partial \mathcal{V}/\partial v_{2,4,5}=0$, we can express VEVs in terms of mass-dimensional parameters of the Higgs potential as follows~\cite{Nomura:2018cfu}:
\begin{align}
v_2\sim \sqrt{\frac{\mu_{H_2}^2}{\lambda_{H_2H_2}} }
, \quad  v_4 \sim \delta\lambda \frac{v_2^3}{\mu^2_4},
\quad  v_5 \sim  v_4\frac{\mu_0}{\mu_5}  \frac{v_2 }{\mu_5},
\end{align}
where $v_2\approx246$ GeV and we abbreviate the lower index of $\delta\lambda_{ij}$. 
While the VEV of the SM Higgs $v_2$ is given by nearly the same form as the SM Higgs potential, the VEVs $v_{4}$ and $v_5$ are generated at one-loop level through $\delta\lambda$.
It suggests that the setup provides a theoretical explanation on the tininess of these large multiplet Higgs VEVs.
Here, it is noteworthy to discuss the cutoff scale on $\delta\lambda$.
 In fact,  $\delta\lambda$ diverges at $r_4=0(\lambda=\infty)$, and we need to determine our relevant energy scale where our model remains valid.

\begin{figure}[tb]
\begin{center}
\includegraphics[width=13cm]{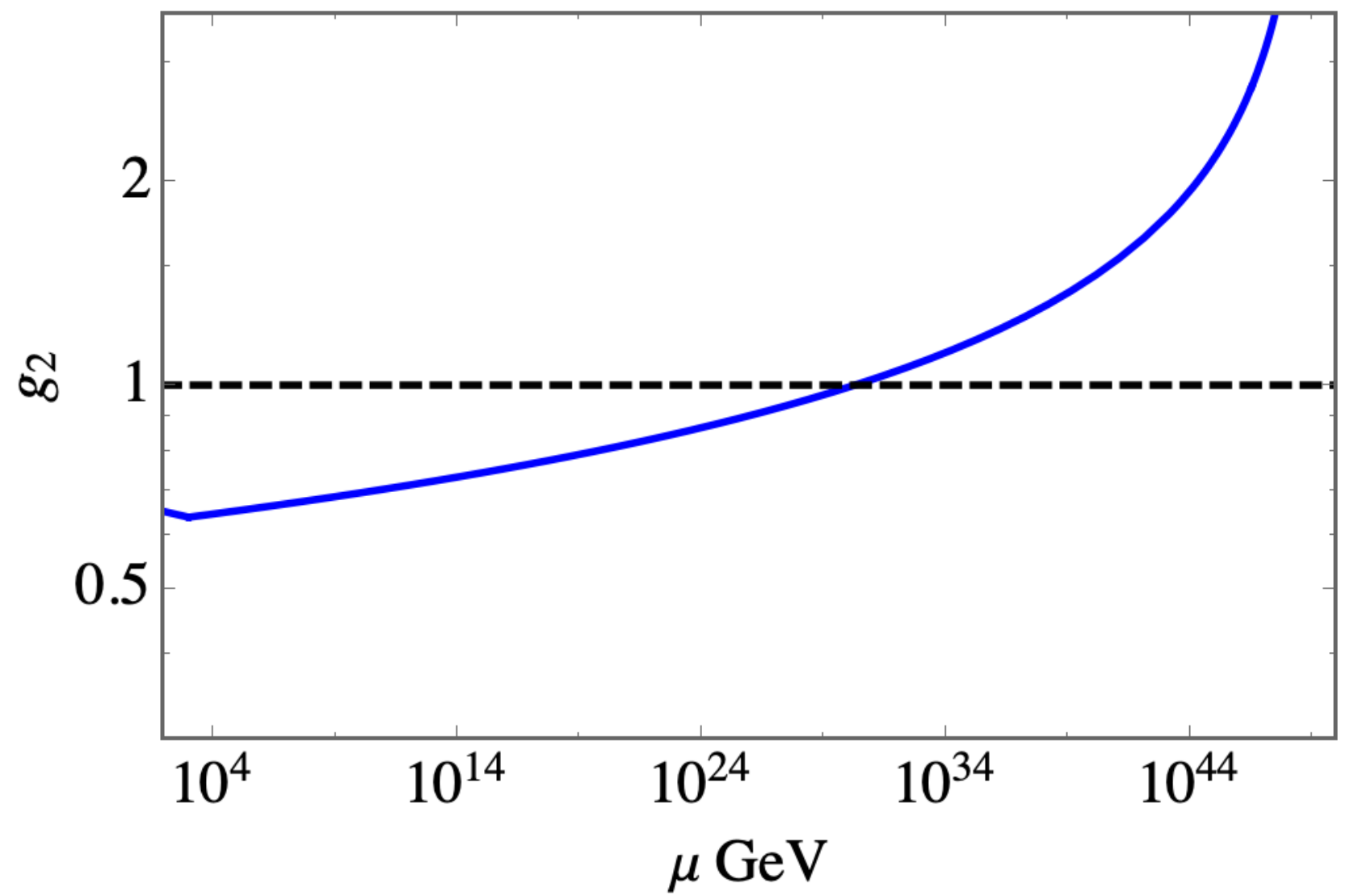}
\caption{The running of $g_2$ in terms of a reference energy of $\mu$, where dashed  horizontal line represents $g_2=1$.}
\label{fig:rge1}
\end{center}\end{figure}

 Since our model introduces two types of large Higgs multiplets,  the SU(2)$_L$ gauge coupling $g_2$ grow rapidly at high energy scale.
 Hence, we fix the cut off scale so that $g_2$ is within the perturbative limit: $g_2\le1$.
 The new beta functions of $g_2$ for quartet Higgs $H_4$ and quintet Higgs $H_5$ are given by
 \begin{align}
 \Delta b^{H_4}_{g_2}=\frac{5}{3} ,  \quad \Delta b^{H_5}_{g_2}=\frac{10}{3} .
\end{align}
Then, the energy evolution of the gauge coupling $g_2$ is~\cite{Kanemura:2015bli}
\begin{align}
\frac{1}{g_2^2(\mu)}&=\frac1{g_2^2(m_{in.})}-\frac{b^{SM}_g}{(4\pi)^2}\ln\left[\frac{\mu^2}{m_{in.}^2}\right]
-\theta(\mu-m_{th.}) \frac{\Delta b^{H_4}_{g_2} + \Delta b^{H_5}_{g_2}}{(4\pi)^2}\ln\left[\frac{\mu^2}{m_{th.}^2}\right],\label{eq:rge_g}
\end{align}
where $\mu$ is a reference energy, $b^{SM}_g=-19/6$, and we set the threshold scale as $m_{th.}=$1000 GeV (with $m_{in.}(=m_Z)< m_{th.}$), which corresponds to the common mass scale of $H_4$ and $H_5$. The $m_{in.}$ and $m_{th.}$ respectively denotes the initial mass and the threshold mass.
The resulting energy flow of $g_2(\mu)$ is shown in Fig.~\ref{fig:rge1}, where dashed horizontal line represents $g_2=1$.
This figure indicates that $g_2$ diverges at around $10^{46}$ GeV, which is much higher than the Planck scale $\sim10^{19}$ GeV.
Thus, any cut-off scale up to Planck scale is relevant to this setup.
When we take some benchmark cut-off energy scales as $(10^5,\ 10^{10},\ 10^{19})$ GeV, assuming $\mu_4=1$ TeV, and $\lambda_0^{(i)} \lambda_{H_2 Hi_4}^{(j)}=0.01$, $ v_4$ is found as
\begin{align}
\Lambda &=10^{19}\ {\rm GeV},\quad  v_4=0.0676\ {\rm GeV}, \label{eq:cut1}\\
\Lambda &=10^{10}\ {\rm GeV},\quad  v_4=0.0285\ {\rm GeV},\label{eq:cut2}\\
\Lambda &=10^{5}\ {\rm GeV},\quad  v_4=0.0068\ {\rm GeV}.\label{eq:cut3}
\end{align}
Thus,  $v_4$ can be successfully suppressed to a level 
two to three orders of magnitude below the experimental bound of a few GeV. 
Moreover, assuming $\mu_0\sim\mu_5\sim1$ TeV, $ v_5$ is further suppressed by approximately one order of magnitude relative to $ v_4$
due to $v_2/\mu_5$.

%

\section{ Applications of our framework}

 In this section, we present several applications of the framework described in the previous section.
The naturally suppressed VEVs of large multiplets are well-suited for explaining the smallness of neutrino masses. 
We then apply our framework to three representative neutrino mass models: model (I) type-III seesaw, model (II) Dirac seesaw and model (III) inverse seesaw.

\subsection{Model (I) : Type-III seesaw}

 \begin{widetext}
\begin{center} 
\begin{table}
\begin{tabular}{|c||c|c|c||c|c|c|}\hline\hline  
&\multicolumn{3}{c||}{Lepton Fields} & \multicolumn{2}{c|}{Scalar Fields} \\\hline
& ~$L_L$~ & ~$e_R$~ & ~$\Sigma_R$ ~ & ~$H_2$~ & ~$H_4$~  \\\hline 
$SU(2)_L$ & $\bm{2}$  & $\bm{1}$  & $\bm{3}$ & $\bm{2}$  & $\bm{4}$  \\\hline 
$U(1)_Y$ & $-\frac12$ & $-1$  & $0$  & $\frac12$ & $\frac12$   \\\hline
${\rm FFR}$ & $\mathbbm{I}$ & $\mathbbm{I}$  & $\tau$  & $\mathbbm{I}$ & $\tau$   \\\hline
\end{tabular}
\caption{Contents of fermion and scalar fields
and their charge assignments under $SU(2)_L\times U(1)_Y$ and Fibonacci fusion rule (FFR), for a type-III seesaw model.}
\label{tab:1}
\end{table}
\end{center}
\end{widetext}

\begin{figure}[tb]
\begin{center}
\includegraphics[width=13cm]{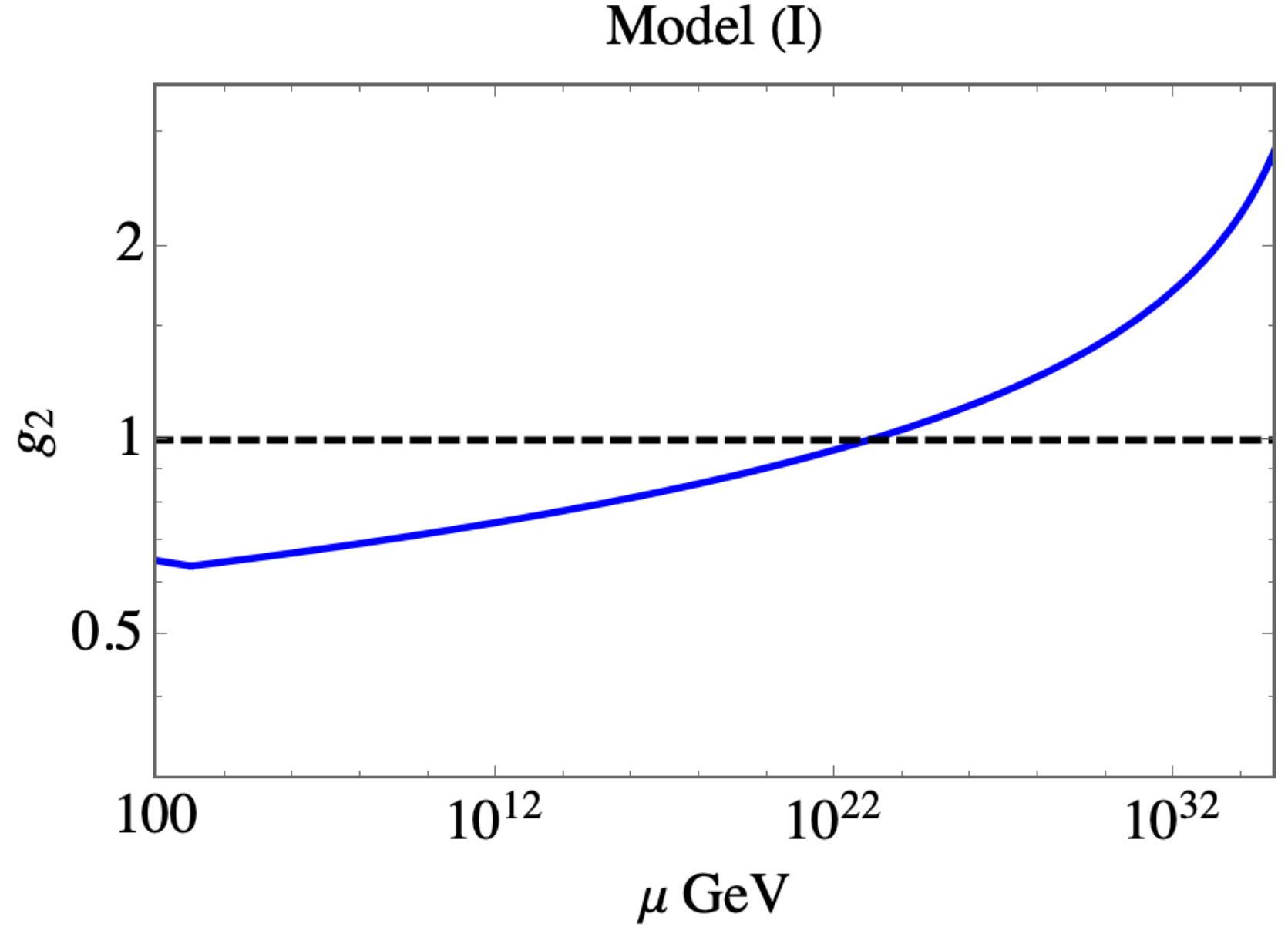}
\caption{The running of $g_2$ in terms of a reference energy of $\mu$, where dashed  horizontal line represents $g_2=1$.}
\label{fig:rge-model1}
\end{center}\end{figure}

 \begin{widetext}
\begin{center} 
\begin{table}
\begin{tabular}{|c||c|c|c||c|}\hline\hline  
&\multicolumn{3}{c||}{$M_R$/GeV}  \\\hline
$\Lambda$/GeV ~& ~$10^2$~ & ~$10^3$~ & ~$10^5$ ~   \\\hline\hline 
$10^{19}$ & $1.48\times10^{-3}$  & $4.68\times10^{-3}$  & $4.68\times10^{-2}$    \\\hline 
$10^{10}$ & $3.51\times10^{-3}$ & $1.11\times10^{-2}$  & $0.111$      \\\hline
$10^{5}$ & $1.47\times10^{-2}$ & $4.65\times10^{-2}$  & $0.465$   \\\hline
\end{tabular}
\caption{Required absolute values of $y_\nu$ for each scale of $\Lambda$ and $M_R$. Here, we suppose the neutrino mass scale is 0.1 eV.}
\label{tab:ynu-1}
\end{table}
\end{center}
\end{widetext}

A type-III model can be achieved by introducing three families of isospin triplet right-handed fermions $\Sigma_R$ which have zero hypercharge.
In addition, we assign $\tau$ to the new particles $\Sigma_R$ and $H_4$ under FFR.
It is notable that $H_5$ is not needed in this model.
 The particle contents and their charges are shown in Tab.~\ref{tab:1}.
 Hereafter, we simply assume the charged-lepton sector is diagonal and will not be considered.
Then, the relevant  Lagrangian under these symmetries is given by
\begin{align}
-\mathcal{L}_{Y}
&=
(y_{\nu})_{ij} [\overline{L_{L_i}} (i\tau_2) H_4^* \Sigma_{R_j} ]
 +  (M_{R})_i [\overline{\Sigma^C_{R_i}} \Sigma_{R_i}] + {\rm h.c.},\label{eq:lep}
\end{align}
where $i=1-3$, $j=1-3$, $\tau_2$ is the second Pauli matrix, and $M_R$ can be taken to be diagonal without loss of generality.
Note that inside bracket "$[ \ ]$" the $SU(2)_L$ indices are contracted so that it makes singlet where we omit details here.
We work on the basis that all the coefficients are real and positive for simplicity.
 Note that the $H_4$ VEV generates Dirac mass term between neutral components of $L_L$ and $\Sigma_R$ in contrast to the original type-III seesaw model where the original term $\overline{L_L} (i\tau_2) H_2^* \Sigma_R$ is forbidden by the FFR.

The neutrino mass matrix $m_\nu$ is given after spontaneous symmetry breaking of $H_4$, and
is given by 
\begin{align}
& (m_{\nu})_{ij}
=v_4^2 \sum_{a=1}^3{(y_\nu)_{ia}M_{R_a}^{-1} (y_\nu^T)_{aj}} .
\end{align}
 The new beta functions of $g_2$ for quartet boson $H_4$ and one triplet $\Sigma_R$ are given by
 \begin{align}
 \Delta b^{H_4}_{g_2}=\frac{5}{3} ,  \quad \Delta b^{\Sigma_R}_{g_2}=\frac{4}{3} .
\end{align}
Then, the energy evolution of the gauge coupling $g_2$ as
\begin{align}
\frac{1}{g_2^2(\mu)} = & \frac1{g_2^2(m_{in.})}-\frac{b^{SM}_g}{(4\pi)^2}\ln\left[\frac{\mu^2}{m_{in.}^2}\right] \nn \\
&-3\theta(\mu-m_{th.}) \frac{\Delta b^{\Sigma_R}_{g_2}}{(4\pi)^2}\ln\left[\frac{\mu^2}{m_{th.}^2}\right] 
-\theta(\mu-m_{th.}) \frac{\Delta b^{H_4}_{g_2}  }{(4\pi)^2}\ln\left[\frac{\mu^2}{m_{th.}^2}\right],\label{eq:rge_g}
\end{align}
where $\mu$ is a reference energy, $b^{SM}_g=-19/6$, and we set the threshold scale as $m_{th.}=$1000 GeV (with $m_{in.}(=m_Z)< m_{th.}$), which corresponds to the common mass scale of $H_4$ and $\Sigma_R$.
The resulting energy flow of $g_2(\mu)$ is shown in Fig.~\ref{fig:rge-model1}, where dashed horizontal line represents $g_2=1$.
This figure indicates that $g_2$ exceeds 1 at around $10^{23}$ GeV, which is much higher than the Planck scale $\sim10^{19}$ GeV.
Thus, we adopt the same benchmark cut-off energy scale in Eqs.~(\ref{eq:cut1})-(\ref{eq:cut3}).
Once we fix three bench mark points $(10^2,\ 10^3,\ 10^5)$ GeV for $M_R$, the required absolute values for $y_\nu$ is listed in Tab.~\ref{tab:ynu-1}.
Here, we suppose the neutrino masses is 0.1 eV and $M_R$ is taken so that all mass scale can reach 
 within the current and future experimental energy that can be realized by the LHC, FCC~\cite{FCC:2018byv,FCC:2025lpp}, CEPC~\cite{CEPCStudyGroup:2023quu,Ai:2025cpj,CEPCStudyGroup:2025kmw} and muon colliders~\cite{Accettura:2023ked,Hamada:2022mua}. 
Obviously, the required Yukawa coupling runs over $10^{-3}$ to $10^{-1}$ orders of magnitude which is  greater than typical seesaw scale.

\subsection{Model (II) : Dirac seesaw}

 \begin{widetext}
\begin{center} 
\begin{table}
\begin{tabular}{|c||c|c|c||c|c|c|c|}\hline\hline  
&\multicolumn{3}{c||}{Lepton Fields} & \multicolumn{3}{c|}{Scalar Fields} \\\hline
& ~$L_L$~ & ~$e_R$~ & ~$\Sigma_{R,L}$ ~ & ~$H_2$~ & ~$H_4$~ & ~$H_5$~  \\\hline 
$SU(2)_L$ & $\bm{2}$  & $\bm{1}$  & $\bm{4}$ & $\bm{2}$& $\bm{4}$  & $\bm{5}$  \\\hline 
$U(1)_Y$ & $-\frac12$ & $-1$  & $\frac12$  & $\frac12$ & $\frac12$ & $1$   \\\hline
${\rm FFR}$ & $\mathbbm{I}$ & $\mathbbm{I}$  & $\tau$  & $\mathbbm{I}$ & $\tau$& $\tau$   \\\hline
\end{tabular}
\caption{Contents of fermion and scalar fields
and their charge assignments under $SU(2)_L\times U(1)_Y$ and Fibonacci fusion rule (FFR), for a Dirac seesaw model.}
\label{tab:2dirac}
\end{table}
\end{center}
\end{widetext}

The Dirac seesaw model can be achieved by introducing three families of isospin quartet vector-like fermions $\Sigma$ which has $1/2$ hypercharge.
In addition, we assign $\tau$ to new particles $\Sigma_{R,L}$, $H_4$, and $H_5$ under FFR, where $\Sigma_L$ is needed in order to cancel the chiral gauge anomalies.
While $H_4$ does not contribute to the Yukawa sector, it is essential for generating the small VEV of $H_5$.
 The particle contents and their charges are shown in Tab.~\ref{tab:2dirac}.
 Hereafter, we simply assume the charged-lepton sector is diagonal and do not consider anymore.
Then, the relevant  Lagrangian under these symmetries is given by
\begin{align}
-\mathcal{L}_{Y}
&=
(y_{\nu})_{ij} [\overline{L_{L_i}} (i\tau_2) H_5^* \Sigma_{R_j} ]
 +  M_{i} [\overline{\Sigma_{L_i}} \Sigma_{R_i}] 
 + {\rm h.c.},\label{eq:lep}
\end{align}
where $i=1-3$, $j=1-3$, $\tau_2$ is the second Pauli matrix, $M$ can be taken to be diagonal without loss of generality.
Note that inside bracket "$[ \ ]$" the $SU(2)_L$ indices are contracted so that it makes singlet where we omit details here.
We work on the basis that all the coefficients are real and positive for simplicity.

 The neutrino mass matrix $m_\nu$ is given after spontaneous symmetry breaking, and
is given by 
\begin{align}
& m_{\nu}
\simeq\frac{v_5}{\sqrt2}y_\nu,
\end{align}
where we neglect subdominant contributions arising from the diagonalization of the neutral lepton mass matrix involving $M_i$.

\begin{figure}[tb]
\begin{center}
\includegraphics[width=13cm]{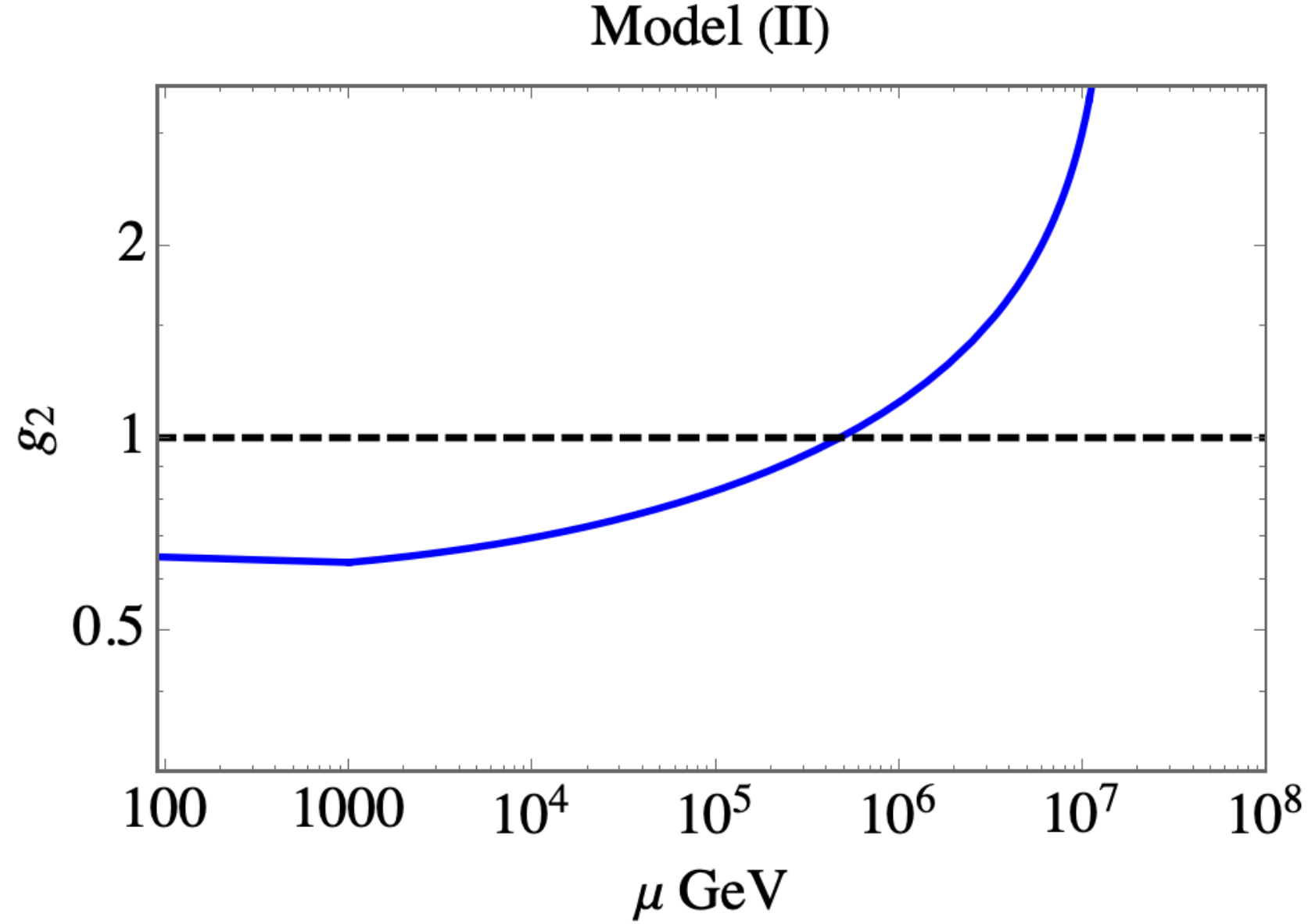}
\caption{The running of $g_2$ in terms of a reference energy of $\mu$, where dashed  horizontal line represents $g_2=1$.}
\label{fig:rge-model2}
\end{center}\end{figure}

 The new beta functions of $g_2$ for quintuplet scalar $H_5$ and one quartet $\Sigma$ are given by
 \begin{align}
 \Delta b^{H_5}_{g_2}=\frac{10}{3} ,  \quad \Delta b^{\Sigma}_{g_2}=\frac{20}{3} .
\end{align}
Then, the energy evolution of the gauge coupling $g_2$ is
\begin{align}
\frac{1}{g_2^2(\mu)} =& \frac1{g_2^2(m_{in.})}-\frac{b^{SM}_g}{(4\pi)^2}\ln\left[\frac{\mu^2}{m_{in.}^2}\right] \nn \\ &
-3 \theta(\mu-m_{th.}) \frac{\Delta b^{\Sigma}_{g_2}}{(4\pi)^2}\ln\left[\frac{\mu^2}{m_{th.}^2}\right]
-\theta(\mu-m_{th.}) \frac{\Delta b^{H_4}_{g_2} +  \Delta b^{H_5}_{g_2} }{(4\pi)^2}\ln\left[\frac{\mu^2}{m_{th.}^2}\right],\label{eq:rge_g}
\end{align}
where $\mu$ is a reference energy, $b^{SM}_g=-19/6$, and we set the threshold scale as $m_{th.}=$1000 GeV (with $m_{in.}(=m_Z)< m_{th.}$), which corresponds to the common mass scale of $H_4$, $H_5$ and $\Sigma$.
The resulting energy flow of $g_2(\mu)$ is shown in Fig.~\ref{fig:rge-model2}, where dashed horizontal line represents $g_2=1$.
This figure indicates that $g_2$ exceeds 1 at around $3\times 10^{5}$ GeV, which implies the valid energy scale in this model.
Thus, we adopt the following benchmark cut-off energy scale $10^{5}$.
Then, with $\delta v_5\approx 6.79\times 10^{-4}$ GeV,  the required Yukawa coupling is 
\begin{align}
|y_\nu|\simeq 2.08\times 10^{-7}.
\end{align}
The value is sufficiently small compared to that in a typical Dirac seesaw model. 
Considering the electron Yukawa coupling is about $2.94\times10^{-6}$, the hierarchy between the charged-lepton and neutrino sectors is acceptably mild.

\subsection{Model (III) : Inverse seesaw}

The inverse seesaw model can be achieved by introducing three families of isospin quartet vector-like fermions $\psi_{L,R}$ and isospin septet right-handed fermions $\Sigma_R$ each of which has $-1/2$ and zero hypercharge~\cite{Nomura:2018cfu}.
In addition, we impose $\tau$ to only new particles $\psi$, $\Sigma_{R}$ and $H_5$ under FFR.
 The particle contents and their charges are shown in Tab.~\ref{tab:3inv}.

 \begin{widetext}
\begin{center} 
\begin{table}
\begin{tabular}{|c||c|c|c|c||c|c|c||c}\hline\hline  
&\multicolumn{4}{c||}{Lepton Fields} & \multicolumn{3}{c|}{Scalar Fields} \\\hline
& ~$L_L$~ & ~$e_R$~ & ~$\psi_{L,R}$ ~ & ~$\Sigma_{R}$ ~ & ~$H_2$~ & ~$H_4$~ & ~$H_5$~  \\\hline 
$SU(2)_L$ & $\bm{2}$  & $\bm{1}$  & $\bm{4}$& $\bm{7}$ & $\bm{2}$  & $\bm{4}$ & $\bm{5}$  \\\hline 
$U(1)_Y$ & $-\frac12$ & $-1$  & $-\frac12$ & $0$  & $\frac12$& $\frac12$ & $1$   \\\hline
${\rm FFR}$ & $\mathbbm{I}$ & $\mathbbm{I}$  & $\tau$  & $\tau$  & $\mathbbm{I}$ & $\tau$ & $\tau$   \\\hline
\end{tabular}
\caption{Contents of fermion and scalar fields
and their charge assignments under $SU(2)_L\times U(1)_Y$ and Fibonacci fusion rule (FFR), for an inverse seesaw model.}
\label{tab:3inv}
\end{table}
\end{center}
\end{widetext}

\begin{figure}[tb]
\begin{center}
\includegraphics[width=13cm]{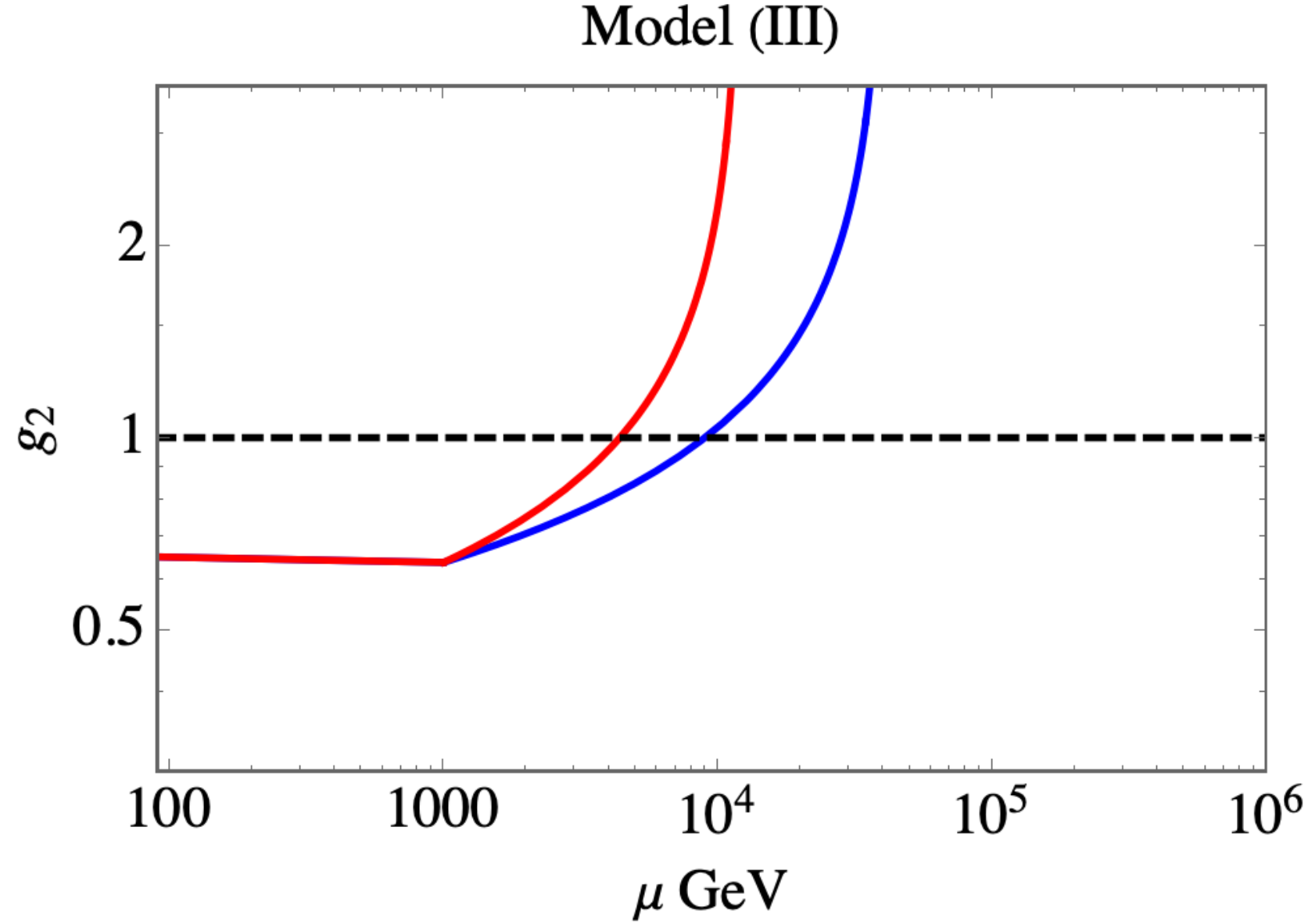}
\caption{The running of $g_2$ in terms of a reference energy of $\mu$, where dashed  horizontal line represents $g_2=1$. The red line is the case of three families of fermions, while the blue line is the case of two families.}
\label{fig:rge-model3}
\end{center}\end{figure}

Then, the relevant  Lagrangian under these symmetries is given by
\begin{align}
-\mathcal{L}_{Y}
&=
(y_{\nu})_{ij} [\overline{L_{L_i}} (i\tau_2) H_5^* \psi^C_{L_j} ]
+f_{ij} [\overline{\psi_{L_i}} (i\tau_2) H_4^* \Sigma_{R_j} ]
+f'_{ij} [\overline{\psi^C_{R_i}}  H_4 \Sigma_{R_j} ] \nn \\
&+M_{i} [\overline{\psi_{L_i}} \psi_{R_i}] 
+M_{\Sigma_i} [\overline{\Sigma^C_{R_i}} \Sigma_{R_i}] + {\rm h.c.},\label{eq:lep}
\end{align}
where $i=1-3$, $j=1-3$, $\tau_2$ is the second Pauli matrix, $M$ and $M_\Sigma$ can be taken to be diagonal without loss of generality.
After the spontaneous symmetry breaking together with integrating out the masses of $\Sigma_R$~\cite{Nomura:2018cfu}, the valid Lagrangian to construct the active neutrino mass matrix via an inverse seesaw is given by
\begin{align}
-\mathcal{L}_{Y}
\sim
m_{D_{ij}} \overline{\nu_{L_i}} \psi^C_{L_j}
+\delta \mu_{R_{ij}} \overline{\psi_{R_i}}  \psi^C_{R_j}
+M_{D_i} \overline{\psi_{L_i}} \psi_{R_i}  + {\rm h.c.},\label{eq:lep}
\end{align}
where $m_D\equiv y_\nu  v_5/\sqrt2$, $\delta \mu_R\equiv v_4^2 f M^{-1}_\Sigma f^T$, and
$\delta\mu_R$ is induced via $f$ and $M$ with analogy of the canonical seesaw.
Then, the neutrino mass matrix $m_\nu$,
which is generated through the inverse seesaw mechanism, 
is given as follows;
\begin{align}
 m_{\nu}
\simeq m^*_D M^{-1}\delta \mu_R^* M^{-1} m_D^T.
\end{align}
 The new beta functions of $g_2$ for quartet boson $H_4$, quintet boson $H_5$  one quartet fermion $\psi$, and one septet fermion $\Sigma_R$ are given by
 \begin{align}
 \Delta b^{H_4}_{g_2}=\frac{5}{3} ,  \quad \Delta b^{\Sigma}_{g_2}=\frac{10}{3} ,
  \Delta b^{\psi}_{g_2}=\frac{20}{3} ,  \quad \Delta b^{\Sigma_R}_{g_2}=\frac{56}{3} .
\end{align}
Then, the energy evolution of the gauge coupling $g_2$ as
\begin{align}
\frac{1}{g_2^2(\mu)}&=\frac1{g_2^2(m_{in.})}-\frac{b^{SM}_g}{(4\pi)^2}\ln\left[\frac{\mu^2}{m_{in.}^2}\right]
-N_f \theta(\mu-m_{th.}) \frac{\Delta b^{\psi}_{g_2}+\Delta b^{\Sigma_R}_{g_2}}{(4\pi)^2}\ln\left[\frac{\mu^2}{m_{th.}^2}\right]\nn\\
&-\theta(\mu-m_{th.}) \frac{\Delta b^{H_4}_{g_2}+ \Delta b^{H_5}_{g_2}}{(4\pi)^2}\ln\left[\frac{\mu^2}{m_{th.}^2}\right],\label{eq:rge_g}
\end{align}
where $N_f$ is the number of $\Psi/\Sigma$ families and the other definitions are the same as the one in the previous models.
The resulting energy flow of $g_2(\mu)$ is shown in Fig.~\ref{fig:rge-model3}, where dashed horizontal line represents $g_2=1$
and the red line is the case of three families of fermions, while the blue line is the case of two families.
This figure indicates that $g_2$ exceeds 1 at around $5$ TeV  (three families of fermions) and 10 TeV (two families of fermions), which implies the valid energy scale in this model.
When we adopt the following benchmark cut-off energy scale $5(10)$ TeV,
we obtain $ v_4 \approx 1.40(2.54)\times 10^{-3}$ GeV.
The neutrino mass matrix is essentially given by
\begin{align}
\frac{( v_4 v_5)^2}{M^2 M_\Sigma} (fy_\nu)^2.
\end{align}
If we adopt the same condition as  the previous condition on $ v_5$; $ v_5\approx   v_4/10$,
the required Yukawa coupling; assuming $f\sim y_\nu$, exceeds 100, which is beyond the perturbative limit.
Therefore, we must consider an alternative scenario.
Here, we assume  $ v_5\approx  v_4\times 10^3$ GeV,
therefore   $\mu_0 v_2/\mu_5^2 \sim 10^3$.
Then, the required Yukawa coupling is marginally within the perturbative limit $\sim4\pi$ as follows:
\begin{align}
|y_\nu|\sim 7 \ {\rm for} \ N_f=2,\quad
|y_\nu|\sim 4\pi \ {\rm for} \ N_f=3,
\end{align}
where we assume $M\sim M_\Sigma\sim1$ TeV.
In this regime, the Yukawa coupling runs faster than $g_2$, causing the model to exit the perturbative regime at lower scales.
To achieve sufficient suppression, the initial Yukawa coupling must be of the order 0.01.
This necessitates $ v_5\approx  v_4\times 10^8$ GeV.
While theoretically possible, this scenario appears unnatural due to the requirement of a large $\mu_0$ parameter.

\section{Summaries and conclusions}
In this paper, we have proposed a novel mechanism to explain the natural smallness of the vacuum expectation values (VEVs) of exotic multi-Higgs fields by utilizing non-invertible symmetries. By imposing the Fibonacci fusion rule (FFR) algebra on the scalar potential, we successfully forbid the VEVs of
$SU(2)_L$ quartet and quintet Higgs fields at the tree level. These VEVs are subsequently generated radiatively at the one-loop level through the breaking of the FFR, providing a more minimal framework than traditional models that require additional loop-inducing particles.

Our numerical analysis demonstrates that for a wide range of cut-off scales—from $10^ 5$ GeV up to the Planck scale—the generated VEV, $v_4$, remains within the range of 0.007 to 0.07 GeV. This is two to three orders of magnitude below the experimental constraints derived from the $\rho$ parameter, which roughly limit such VEVs to the GeV scale.

Furthermore, we applied this mechanism to three concrete neutrino mass frameworks:
\begin{itemize}
\item  Model (I) Type-III Seesaw: This model remains perturbative up to the Planck scale, requiring Yukawa couplings of $10^{-3}$  to $10^{-1}$ to achieve the observed neutrino mass scale of 0.1 eV.
\item  Model (II) Dirac Seesaw: The gauge coupling $g_2$ reaches the perturbative limit at approximately $3\times10^5$ GeV, allowing for a milder hierarchy between the charged-lepton and neutrino sectors compared to typical tree-level Dirac models.
\item Model (III) Inverse Seesaw: Due to the introduction of larger multiplets, the running of $g_2$ is significantly accelerated, limiting the valid energy scale of the model to approximately 5–10 TeV. While theoretically possible, achieving the correct neutrino mass in this framework may require a relatively large
$\mu_0$ parameter to maintain perturbativity of the Yukawa couplings
\end{itemize}

 We have found these models are highly testable, assuming a new physics scale at the TeV level.
They could be tested in future experiments such as the high-luminosity LHC, ILC, FCC, CEPC and muon colliders.

In conclusion, the integration of non-invertible symmetries into Higgs physics provides a robust theoretical foundation for the existence of small VEVs in extended scalar multiplets. This approach not only ensures consistency with precision electroweak data but also opens new avenues for constructing minimal and testable frameworks, including neutrino mass models and lepton seesaw scenarios.

\section*{Acknowledgments}
\vspace{0.5cm}
TN is supported by the Fundamental Research Funds for the Central Universities.
HO is supported by Zhongyuan Talent (Talent Recruitment Series) Foreign Experts Project.
\bibliography{references.bib}

@article{Xu:2026nwh,
    author = "Xu, Ling-Xiao",
    title = "{A General Prescription for Spurion Analysis of Non-Invertible Selection Rules}",
    eprint = "2604.09345",
    archivePrefix = "arXiv",
    primaryClass = "hep-ph",
    month = "4",
    year = "2026"
}

@article{Okada:2026bpp,
    author = "Okada, Hiroshi and Otsuka, Hajime",
    title = "{Dynamical CP Violation from Non-Invertible Selection Rules}",
    eprint = "2604.04423",
    archivePrefix = "arXiv",
    primaryClass = "hep-ph",
    reportNumber = "KYUSHU-HET-353",
    month = "4",
    year = "2026"
}

@article{Okada:2026iob,
    author = "Okada, Hiroshi and Wu, Jia-Jun",
    title = "{Dynamical Determination of the Cut-off Scale in Loop-Induced Neutrino Mass Models with Non-Invertible Symmetry}",
    eprint = "2603.17587",
    archivePrefix = "arXiv",
    primaryClass = "hep-ph",
    month = "3",
    year = "2026"
}

@article{Okada:2026pek,
    author = "Okada, Hiroshi and Singh, Labh",
    title = "{Dirac one-loop seesaw in a non-invertible fusion rule}",
    eprint = "2604.11308",
    archivePrefix = "arXiv",
    primaryClass = "hep-ph",
    month = "4",
    year = "2026"
}

@article{Kanemura:2011mw,
    author = "Kanemura, Shinya and Nabeshima, Takehiro and Sugiyama, Hiroaki",
    title = "{TeV-Scale Seesaw with Loop-Induced Dirac Mass Term and Dark Matter from $U(1)_{B-L}$ Gauge Symmetry Breaking}",
    eprint = "1111.0599",
    archivePrefix = "arXiv",
    primaryClass = "hep-ph",
    reportNumber = "UT-HET-060",
    doi = "10.1103/PhysRevD.85.033004",
    journal = "Phys. Rev. D",
    volume = "85",
    pages = "033004",
    year = "2012"
}

@article{Nomura:2026hli,
    author = "Nomura, Takaaki and Okada, Hiroshi and Shigekami, Yoshihiro",
    title = "{A Minimal Realization of Radiative Dirac Neutrino Masses via a Non-Invertible Fusion Rule}",
    eprint = "2603.15382",
    archivePrefix = "arXiv",
    primaryClass = "hep-ph",
    month = "3",
    year = "2026"
}

@article{Kanemura:2015bli,
    author = "Kanemura, Shinya and Nishiwaki, Kenji and Okada, Hiroshi and Orikasa, Yuta and Park, Seong Chan and Watanabe, Ryoutaro",
    title = "{LHC 750 GeV diphoton excess in a radiative seesaw model}",
    eprint = "1512.09048",
    archivePrefix = "arXiv",
    primaryClass = "hep-ph",
    reportNumber = "CTPU-15-29, KIAS-P15067, UT-HET-110",
    doi = "10.1093/ptep/ptw164",
    journal = "PTEP",
    volume = "2016",
    number = "12",
    pages = "123B04",
    year = "2016"
}

@article{Nomura:2018cfu,
    author = "Nomura, Takaaki and Okada, Hiroshi",
    title = "{Inverse seesaw model with a natural hierarchy at the TeV scale}",
    eprint = "1807.04555",
    archivePrefix = "arXiv",
    primaryClass = "hep-ph",
    reportNumber = "KIAS-P18078, APCTP Pre2018 - 007",
    doi = "10.1103/PhysRevD.99.055027",
    journal = "Phys. Rev. D",
    volume = "99",
    number = "5",
    pages = "055027",
    year = "2019"
}

@article{ParticleDataGroup:2024cfk,
    author = "Navas, S. and others",
    collaboration = "Particle Data Group",
    title = "{Review of particle physics}",
    doi = "10.1103/PhysRevD.110.030001",
    journal = "Phys. Rev. D",
    volume = "110",
    number = "3",
    pages = "030001",
    year = "2024"
}

@article{CEPCStudyGroup:2023quu,
    author = "Abdallah, Waleed and others",
    collaboration = "CEPC Study Group",
    title = "{CEPC Technical Design Report: Accelerator}",
    eprint = "2312.14363",
    archivePrefix = "arXiv",
    primaryClass = "physics.acc-ph",
    reportNumber = "IHEP-CEPC-DR-2023-01, IHEP-AC-2023-01",
    doi = "10.1007/s41605-024-00463-y",
    journal = "Radiat. Detect. Technol. Methods",
    volume = "8",
    number = "1",
    pages = "1--1105",
    year = "2024",
    note = "[Erratum: Radiat.Detect.Technol.Methods 9, 184--192 (2025)]"
}

@article{Nakai:2025thw,
    author = "Nakai, Yuichiro and Otsuka, Hajime and Shigekami, Yoshihiro and Zhang, Zhihao",
    title = "{The Minimal Supersymmetric Standard Model with Non-Invertible Selection Rules}",
    eprint = "2512.21509",
    archivePrefix = "arXiv",
    primaryClass = "hep-ph",
    reportNumber = "KYUSHU-HET-347",
    month = "12",
    year = "2025"
}

@article{Qu:2026omn,
    author = "Qu, Bu-Yao and Jiang, Zheng and Ding, Gui-Jun",
    title = "{Two-zero textures of the Majorana neutrino mass matrix from $\mathbb{Z}_3$ gauging of $\mathbb{Z}_N$ non-invertible symmetry}",
    eprint = "2602.24214",
    archivePrefix = "arXiv",
    primaryClass = "hep-ph",
    month = "2",
    year = "2026"
}

@article{Kanemura:2012rj,
    author = "Kanemura, Shinya and Sugiyama, Hiroaki",
    title = "{Dark matter and a suppression mechanism for neutrino masses in the Higgs triplet model}",
    eprint = "1202.5231",
    archivePrefix = "arXiv",
    primaryClass = "hep-ph",
    reportNumber = "UT-HET-065",
    doi = "10.1103/PhysRevD.86.073006",
    journal = "Phys. Rev. D",
    volume = "86",
    pages = "073006",
    year = "2012"
}

@article{Kobayashi:2025thd,
    author = "Kobayashi, Tatsuo and Otsuka, Hajime and Yanagida, Tsutomu T.",
    title = "{Non-invertible Symmetry as a Solution to the Strong CP Problem in a GUT-inspired Standard Model}",
    eprint = "2508.12287",
    archivePrefix = "arXiv",
    primaryClass = "hep-ph",
    reportNumber = "EPHOU-25-014, KYUSHU-HET-334",
    month = "8",
    year = "2025"
}

@article{Kobayashi:2025rpx,
    author = "Kobayashi, Tatsuo and Otsuka, Hajime and Tanimoto, Morimitsu and Yanagida, Tsutomu T.",
    title = "{GUT-motivated non-invertible symmetry as a solution to the strong CP problem and the neutrino CP-violating phase}",
    eprint = "2510.01680",
    archivePrefix = "arXiv",
    primaryClass = "hep-ph",
    reportNumber = "EPHOU-25-018, KYUSHU-HET-340",
    month = "10",
    year = "2025"
}

@article{Okada:2025kfm,
    author = "Okada, Hiroshi and Shigekami, Yoshihiro",
    title = "{Three-loop induced neutrino mass model in a non-invertible symmetry}",
    eprint = "2507.16198",
    archivePrefix = "arXiv",
    primaryClass = "hep-ph",
    month = "7",
    year = "2025"
}

@article{Chen:2025awz,
    author = "Chen, Jingqian and Geng, Chao-Qiang and Okada, Hiroshi and Wu, Jia-Jun",
    title = "{A radiative lepton model in a non-invertible fusion rule}",
    eprint = "2507.11951",
    archivePrefix = "arXiv",
    primaryClass = "hep-ph",
    month = "7",
    year = "2025"
}

@article{Jangid:2025krp,
    author = "Jangid, Shilpa and Okada, Hiroshi",
    title = "{A natural realization of inverse seesaw model in a non-invertible selection rule}",
    eprint = "2508.16174",
    archivePrefix = "arXiv",
    primaryClass = "hep-ph",
    month = "8",
    year = "2025"
}

@article{Jangid:2025thp,
    author = "Jangid, Shilpa and Okada, Hiroshi",
    title = "{A radiative seesaw model in a non-invertible selection rule with the assistance of a non-holomorphic modular $A_4$ symmetry}",
    eprint = "2510.17292",
    archivePrefix = "arXiv",
    primaryClass = "hep-ph",
    month = "10",
    year = "2025"
}

@article{Nomura:2025sod,
    author = "Nomura, Takaaki and Okada, Hiroshi",
    title = "{Radiative lepton seesaw model in a non-invertible fusion rule and gauged $B-L$ symmetry}",
    eprint = "2506.16706",
    archivePrefix = "arXiv",
    primaryClass = "hep-ph",
    month = "6",
    year = "2025"
}

@article{Nomura:2025tvz,
    author = "Nomura, Takaaki and Okada, Hiroshi and Shigekami, Yoshihiro",
    title = "{Radiative lepton model in a non-invertible fusion rule}",
    eprint = "2510.17156",
    archivePrefix = "arXiv",
    primaryClass = "hep-ph",
    month = "10",
    year = "2025"
}

@article{Okada:2026gxl,
    author = "Okada, Hiroshi and Shigekami, Yoshihiro",
    title = "{Two-loop rainbow neutrino masses in a non-invertible symmetry}",
    eprint = "2601.15749",
    archivePrefix = "arXiv",
    primaryClass = "hep-ph",
    month = "1",
    year = "2026"
}

@article{Okada:2025adm,
    author = "Okada, Hiroshi and Shoji, Yutaro",
    title = "{A novel realization of linear seesaw model in a non-invertible selection rule with the assistance of $\mathbb Z_3$ symmetry}",
    eprint = "2512.20891",
    archivePrefix = "arXiv",
    primaryClass = "hep-ph",
    month = "12",
    year = "2025"
}

@article{Kobayashi:2025wty,
    author = "Kobayashi, Tatsuo and Otsuka, Hajime",
    title = "{Generalized CP from non-invertible selection rules}",
    eprint = "2512.16376",
    archivePrefix = "arXiv",
    primaryClass = "hep-ph",
    reportNumber = "EPHOU-25-019,KYUSHU-HET-345",
    month = "12",
    year = "2025"
}

@article{Nomura:2025yoa,
    author = "Nomura, Takaaki and Popov, Oleg",
    title = "{No-group Scotogenic Model}",
    eprint = "2507.10299",
    archivePrefix = "arXiv",
    primaryClass = "hep-ph",
    month = "7",
    year = "2025"
}

@article{Kobayashi:2024cvp,
    author = "Kobayashi, Tatsuo and Otsuka, Hajime and Tanimoto, Morimitsu",
    title = "{Yukawa textures from non-invertible symmetries}",
    eprint = "2409.05270",
    archivePrefix = "arXiv",
    primaryClass = "hep-ph",
    reportNumber = "EPHOU-24-013, KYUSHU-HET-295",
    doi = "10.1007/JHEP12(2024)117",
    journal = "JHEP",
    volume = "12",
    pages = "117",
    year = "2024"
}

@article{Kobayashi:2025znw,
    author = "Kobayashi, Tatsuo and Nishioka, Yume and Otsuka, Hajime and Tanimoto, Morimitsu",
    title = "{More about quark Yukawa textures from selection rules without group actions}",
    eprint = "2503.09966",
    archivePrefix = "arXiv",
    primaryClass = "hep-ph",
    reportNumber = "EPHOU-25-003, KYUSHU-HET-312",
    doi = "10.1007/JHEP05(2025)177",
    journal = "JHEP",
    volume = "05",
    pages = "177",
    year = "2025"
}

@article{Suzuki:2025oov,
    author = "Suzuki, Motoo and Xu, Ling-Xiao",
    title = "{Phenomenological implications of a class of non-invertible selection rules}",
    eprint = "2503.19964",
    archivePrefix = "arXiv",
    primaryClass = "hep-ph",
    month = "3",
    year = "2025"
}

@article{Liang:2025dkm,
    author = "Liang, Qiuyue and Yanagida, Tsutomu T.",
    title = "{Non-invertible symmetry as an axion-less solution to the strong CP problem}",
    eprint = "2505.05142",
    archivePrefix = "arXiv",
    primaryClass = "hep-ph",
    month = "5",
    year = "2025"
}

@article{Kobayashi:2025ldi,
    author = "Kobayashi, Tatsuo and Otsuka, Hajime and Tanimoto, Morimitsu and Uchida, Haruki",
    title = "{Lepton mass textures from non-invertible multiplication rules}",
    eprint = "2505.07262",
    archivePrefix = "arXiv",
    primaryClass = "hep-ph",
    reportNumber = "EPHOU-25-007, KYUSHU-HET-321",
    month = "5",
    year = "2025"
}

@article{Heckman:2024obe,
    author = "Heckman, Jonathan J. and McNamara, Jacob and Montero, Miguel and Sharon, Adar and Vafa, Cumrun and Valenzuela, Irene",
    title = "{On the Fate of Stringy Non-Invertible Symmetries}",
    eprint = "2402.00118",
    archivePrefix = "arXiv",
    primaryClass = "hep-th",
    month = "1",
    year = "2024"
}

@article{Kaidi:2024wio,
    author = "Kaidi, Justin and Tachikawa, Yuji and Zhang, Hao Y.",
    title = "{On a class of selection rules without group actions in field theory and string theory}",
    eprint = "2402.00105",
    archivePrefix = "arXiv",
    primaryClass = "hep-th",
    month = "1",
    year = "2024"
}

@article{Funakoshi:2024uvy,
    author = "Funakoshi, Shuta and Kobayashi, Tatsuo and Otsuka, Hajime",
    title = "{Quantum aspects of non-invertible flavor symmetries in intersecting/magnetized D-brane models}",
    eprint = "2412.12524",
    archivePrefix = "arXiv",
    primaryClass = "hep-th",
    reportNumber = "EPHOU-24-017, KYUSHU-HET-304",
    month = "12",
    year = "2024"
}

@article{Kobayashi:2025lar,
    author = "Kobayashi, Tatsuo and Mita, Hironobu and Otsuka, Hajime and Sakuma, Riku",
    title = "{Matter symmetries in supersymmetric standard models from non-invertible selection rules}",
    eprint = "2506.10241",
    archivePrefix = "arXiv",
    primaryClass = "hep-ph",
    reportNumber = "EPHOU-25-009, KYUSHU-HET-325",
    month = "6",
    year = "2025"
}

@article{Kobayashi:2025cwx,
    author = "Kobayashi, Tatsuo and Okada, Hiroshi and Otsuka, Hajime",
    title = "{Radiative neutrino mass models from non-invertible selection rules}",
    eprint = "2505.14878",
    archivePrefix = "arXiv",
    primaryClass = "hep-ph",
    reportNumber = "EPHOU-25-008, KYUSHU-HET-324",
    month = "5",
    year = "2025"
}

@article{FCC:2025lpp,
    author = "Benedikt, M. and others",
    collaboration = "FCC",
    title = "{Future Circular Collider Feasibility Study Report: Volume 1, Physics, Experiments, Detectors}",
    eprint = "2505.00272",
    archivePrefix = "arXiv",
    primaryClass = "hep-ex",
    reportNumber = "CERN-FCC-PHYS-2025-0002",
    doi = "10.1140/epjc/s10052-025-15077-x",
    journal = "Eur. Phys. J. C",
    volume = "85",
    number = "12",
    pages = "1468",
    year = "2025"
}

@article{FCC:2018byv,
    author = "Abada, A. and others",
    collaboration = "FCC",
    title = "{FCC Physics Opportunities}: {Future Circular Collider Conceptual Design Report Volume 1}",
    reportNumber = "CERN-ACC-2018-0056",
    doi = "10.1140/epjc/s10052-019-6904-3",
    journal = "Eur. Phys. J. C",
    volume = "79",
    number = "6",
    pages = "474",
    year = "2019"
}

@article{CEPCStudyGroup:2025kmw,
    author = "Adhya, Souvik Priyam and others",
    collaboration = "CEPC Study Group",
    title = "{CEPC Technical Design Report - Reference Detector}",
    eprint = "2510.05260",
    archivePrefix = "arXiv",
    primaryClass = "hep-ex",
    reportNumber = "IHEP-CEPC-DR-2025-01, IHEP-EP-2025-01",
    month = "10",
    year = "2025"
}

@article{Hamada:2022mua,
    author = "Hamada, Yu and Kitano, Ryuichiro and Matsudo, Ryutaro and Takaura, Hiromasa and Yoshida, Mitsuhiro",
    title = "{$\mu$TRISTAN}",
    eprint = "2201.06664",
    archivePrefix = "arXiv",
    primaryClass = "hep-ph",
    reportNumber = "KEK-TH-2385",
    doi = "10.1093/ptep/ptac059",
    journal = "PTEP",
    volume = "2022",
    number = "5",
    pages = "053B02",
    year = "2022"
}

@article{Ai:2025cpj,
    author = "Ai, Xiaocong and others",
    title = "{New physics search at the CEPC: a general perspective}",
    eprint = "2505.24810",
    archivePrefix = "arXiv",
    primaryClass = "hep-ex",
    doi = "10.1088/1674-1137/ae1194",
    journal = "Chin. Phys. C",
    volume = "49",
    pages = "123108",
    year = "2025"
}

@article{Accettura:2023ked,
    author = "Accettura, Carlotta and others",
    title = "{Towards a muon collider}",
    eprint = "2303.08533",
    archivePrefix = "arXiv",
    primaryClass = "physics.acc-ph",
    reportNumber = "FERMILAB-PUB-23-123-AD-PPD-T",
    doi = "10.1140/epjc/s10052-023-11889-x",
    journal = "Eur. Phys. J. C",
    volume = "83",
    number = "9",
    pages = "864",
    year = "2023",
    note = "[Erratum: Eur.Phys.J.C 84, 36 (2024)]"
}
\end{document}